\documentclass[twocolumn, switch]{article} 
\usepackage[braket, qm]{qcircuit}
\usepackage{preprint}
\usepackage[export]{adjustbox}
\usepackage{amsmath, amsthm, amssymb, amsfonts}
\usepackage{comment}
\usepackage[numbers,square]{natbib}
\usepackage[utf8]{inputenc}	
\usepackage[T1]{fontenc}	
\usepackage{xcolor}		
\usepackage[colorlinks = true,
            linkcolor = purple,
            urlcolor  = blue,
            citecolor = cyan,
            anchorcolor = black]{hyperref}	
\usepackage{booktabs} 		
\usepackage{nicefrac}		
\usepackage{microtype}		
\usepackage{lineno}		
\usepackage{float}			
\usepackage{algorithm}
\usepackage{algpseudocode}

\usepackage{lipsum}		
\usepackage{subfig}
\usepackage{newfloat}
\DeclareFloatingEnvironment[name={Supplementary Figure}]{suppfigure}
\usepackage{sidecap}
\sidecaptionvpos{figure}{c}

\usepackage{titlesec}
\titlespacing\section{0pt}{12pt plus 3pt minus 3pt}{1pt plus 1pt minus 1pt}
\titlespacing\subsection{0pt}{10pt plus 3pt minus 3pt}{1pt plus 1pt minus 1pt}
\titlespacing\subsubsection{0pt}{8pt plus 3pt minus 3pt}{1pt plus 1pt minus 1pt}

\usepackage{tikz,xcolor,hyperref}

\definecolor{lime}{HTML}{A6CE39}
\DeclareRobustCommand{\orcidicon}{
	\begin{tikzpicture}
	\draw[lime, fill=lime] (0,0) 
	circle [radius=0.16] 
	node[white] {{\fontfamily{qag}\selectfont \tiny ID}};
	\draw[white, fill=white] (-0.0625,0.095) 
	circle [radius=0.007];
	\end{tikzpicture}
	\hspace{-2mm}
}
\foreach \x in {A, ..., Z}{\expandafter\xdef\csname orcid\x\endcsname{\noexpand\href{https://orcid.org/\csname orcidauthor\x\endcsname}
			{\noexpand\orcidicon}}
}

\title{Qurzon: A Prototype for a Divide and Conquer Based Quantum Compiler for Distributed Quantum Systems}

\usepackage{xwatermark}
\newwatermark[firstpage,color=gray!60,angle=90,scale=0.32, xpos=-4.05in,ypos=0]{\href{https://doi.org/}{\color{gray}{}}}
\newwatermark[firstpage,color=gray!60,angle=90,scale=0.32, xpos=3.9in,ypos=0]{\href{https://doi.org/}{\color{gray}{}}}

\author{
	{\hspace{1mm}Turbasu Chatterjee} \\
	Maulana Abul Kalam Azad University of Technology\\
	\texttt{turbasu.chatterjee@gmail.com} \\
	\And
    {\hspace{0mm}Arnav Das} \\
	Kazi Nazrul University\\
	\texttt{arnav.das88@gmail.com} \\
	\And
	{\hspace{1mm}Shah Ishmam Mohtashim} \\
	University of Dhaka\\
	\texttt{sishmam51@gmail.com} \\
	\And
	{\hspace{1mm}Amit Saha} \\
	Atos\\
	\texttt{abamitsaha@gmail.com} \\
 	\And
 	{\hspace{1mm}Amlan Chakrabarti} \\
 	University of Calcutta\\
 	\texttt{acakcs@caluniv.ac.in} \\
}

\begin{document}

\twocolumn[ 
  \begin{@twocolumnfalse} 
  
\maketitle

\begin{abstract}
 
\end{abstract}
\textbf{When working with algorithms on quantum devices, quantum memory becomes a crucial bottleneck due to low qubit count in NISQ-era devices. In this context, the concept of `divide and compute', wherein a quantum circuit is broken into several subcircuits and executed separately, while stitching the results of the circuits via classical post-processing, becomes a viable option, especially in NISQ-era devices. This paper introduces \textbf{Qurzon}, a proposed novel quantum compiler that incorporates the marriage of techniques of divide and compute with the state-of-the-art algorithms of optimal qubit placement for executing on real quantum devices. A scheduling algorithm is also introduced within the compiler that can explore the power of distributed quantum computing while paving the way for quantum parallelism for large algorithms. Several benchmark circuits have been executed using the compiler, thereby demonstrating the power of the divide and compute when working with real NISQ-era quantum devices.}

\vspace{0.35cm}

  \end{@twocolumnfalse} 
] 



\section{Introduction}

Quantum Computing is the future promise of computing with far-reaching impacts in diverse scientific fields \cite{nielsen_chuang_2010} \cite{preskill2018quantum} \cite{steane1998quantum} \cite{williams2010explorations}. It should deliver speed up in many problems of those fields ranging from machine learning \cite{biamonte2017quantum} \cite{schuld2015introduction} to natural sciences \cite{cao2019quantum}. These absolute speedups will occur when large-scale, universal, fault-tolerant quantum computers come into being. Till then, the challenge of the research community is to utilize currently available noisy intermediate-scale quantum computers (NISQ) era devices to their full potential \cite{preskill2018quantum}. These NISQ-era devices are bound by noise levels, limited coherence times of the qubits, scalability, etc \cite{dasgupta2020characterizing}. Thus, researchers are bound to only using small-scale quantum computers with limited and differing connectivity among qubits of different existing quantum devices. 

For this reason, this paper proposes \textbf{Qurzon}: a divide and conquer based quantum compiler. The proposed quantum compiler attempts to make many exponential size problems tractable using current NISQ-era quantum computers by cutting the quantum circuits for those problems to subcircuits that fit the present device architectures. Extending that notion, \textbf{Qurzon} also implements optimal qubit routing for these NISQ-era devices. The nature of NISQ-era devices is that each quantum device is characterized by its inherent qubit connectivity or so-called qubit topology. \textbf{Qurzon} considers these qubit topologies and implements an optimal qubit routing scheme for the subcircuits formed by the cut algorithm.

For this paper, the engineering of \textbf{Qurzon}, a combination of recursive circuit cutting and optimal qubit routing \cite{cowtan2019qubit}, is explained and implemented against existing circuit optimization methods. A significant quantitative advantage is shown that makes \textbf{Qurzon} the state-of-the-art compiler amongst all existing ones found in literature \cite{Ferrari_2021} \cite{ferrarisoftware} \cite{Cuomo_2020} \cite{gyongyosi2021distributed}, thus improving on, and paving the way for future work on quantum multiprocessor computing. 

The key contributions of this paper are as follows: 
\begin{itemize}
    \item Proposing the use of the divide-and-compute framework for execution on actual quantum devices. 
    \item Proposal for a scheduling algorithm for distributed parallel quantum multiprocessing.
    \item Demonstrating the use of optimal qubit routing and evaluating its performance for implementation on actual quantum devices.
    \item Runtime evaluation for the parallel and distributed execution of subcircuits on quantum devices.
\end{itemize}

The organization of the paper is as follows. A few preliminaries of NISQ-era quantum computation are presented in Section 2. Section 3 illustrates the overview of the proposed compiler. In Section 4, experimental results are exhibited with brief discussion. Section 5 depicts our concluding remarks with future scope.

\section{Background}

The circuit-based quantum computing model is one  in which operations are performed by a series of quantum gates that are always reversible. The quantum gates act on the prepared initial qubit states sequentially, transforming the initial states through unitary transformation into the final state as dictated by the quantum operations. The final state is then measured to complete the computation procedure. This reversibility property makes it unique from classical computing. All quantum gates are by-design unitary to ensure reversibility.

Currently, the world has functional but small scale and imperfect quantum computers that are called Noisy intermediate scale Quantum Computers (NISQ). Large-scale quantum computation is impossible in NISQ-era quantum computers due to the increase of errors as the number of qubits is scaled up.  The error propagates throughout the circuits as the number of qubits increases till it reaches a point when measurement on quantum computing becomes hugely unreliable to the point of being untenable.  The current research trend is to improve the existing faulty NISQ devices so that they become viable for usage in the passing time.  Till the era of fault-tolerant computing starts, NISQ devices are the only machines to test and build quantum algorithms. 

Universal fault-tolerant quantum computing (FTQC) \cite{steane1998quantum} \cite{preskill1998fault} \cite{shor1996fault} still requires years of research till they are realizable in real life. FTQCs are large-scale quantum computers with a minimal error rate and significant coherent time, ensuring a significant quantum advantage over classical computing when they arrive. Till then, NISQ devices are currently the best working model of quantum computing we have, and all quantum algorithms are designed keeping the limitations of NISQ in mind. NISQ-era computers are not changing the world right away, but it is a promise of the quantum advantage that is coming. These quantum computers have already some real-life applications that give somewhat exploratory results for some problems \cite{Shor_1997} \cite{Harrow_2009} \cite{mi2021observation} and a few quantum supremacy \cite{arute2019quantum} \cite{harrow2017quantum} \cite{bremner2017achieving} examples have also been shown in recent years. 


NISQ devices are made of qubits that are erroneous, unstable, and have prepared states that decay over short periods \cite{dasgupta2020characterizing} \cite{9390130}. Even the gate operations are faulty and prone to deviations from actual results. Qubit stability error is called decoherence whereas gate operation error is known as fidelity.  Due to these errors, the quantum computing hardware cannot scale. Hence, the term intermediate scale came into being. Each of these NISQ-era devices have particular topologies or connectivities that acts as a characterization of these devices.

\begin{figure*}[h!]
\begin{center}
    \begin{tabular}{c}
        \subfloat[]{\includegraphics[width = 3.0in]{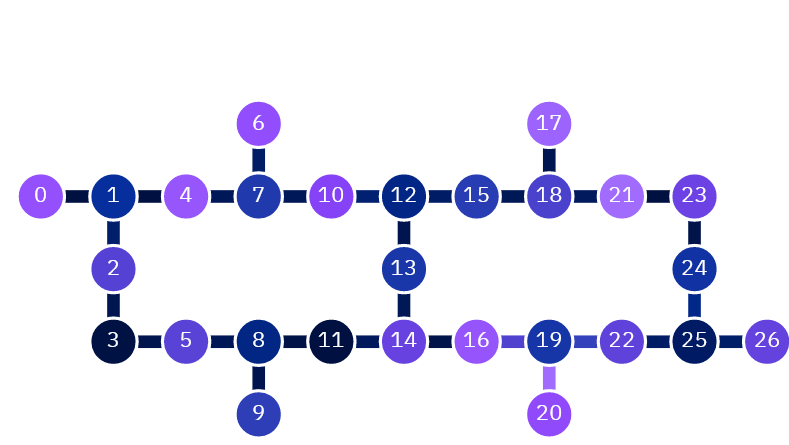}}
  \end{tabular}
  \begin{tabular}{cc}
    \subfloat[]{\includegraphics[width=2.0in]{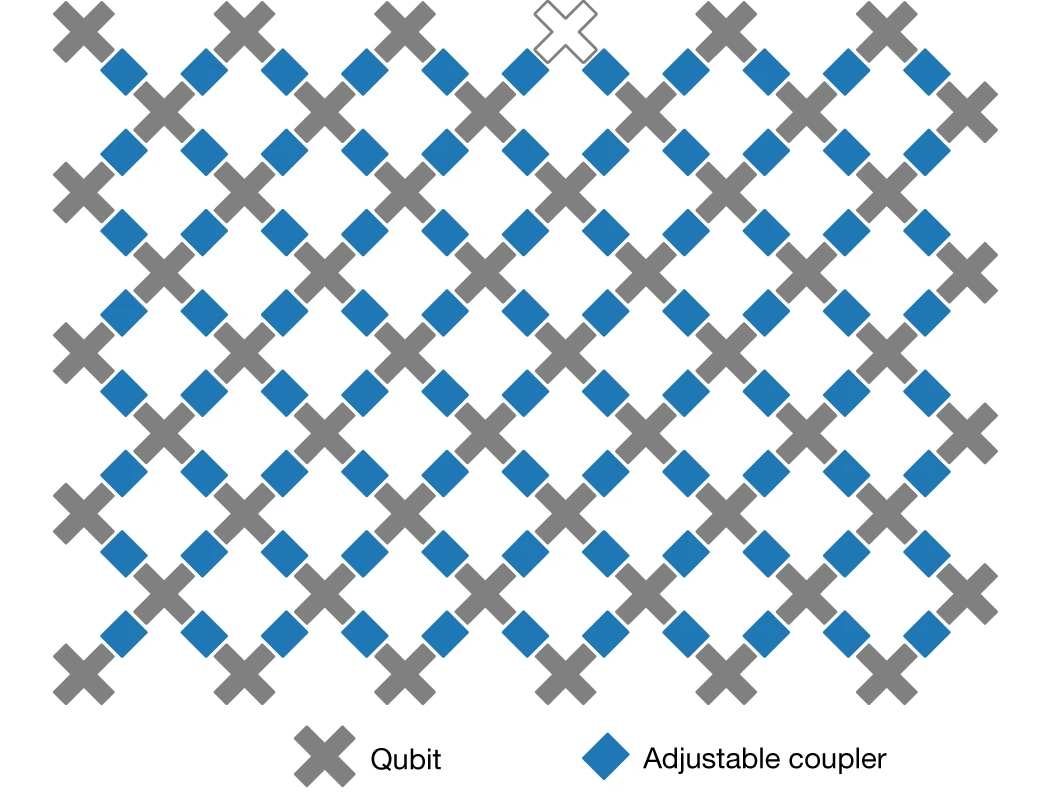}} &
    \subfloat[]{\includegraphics[width=1.5in]{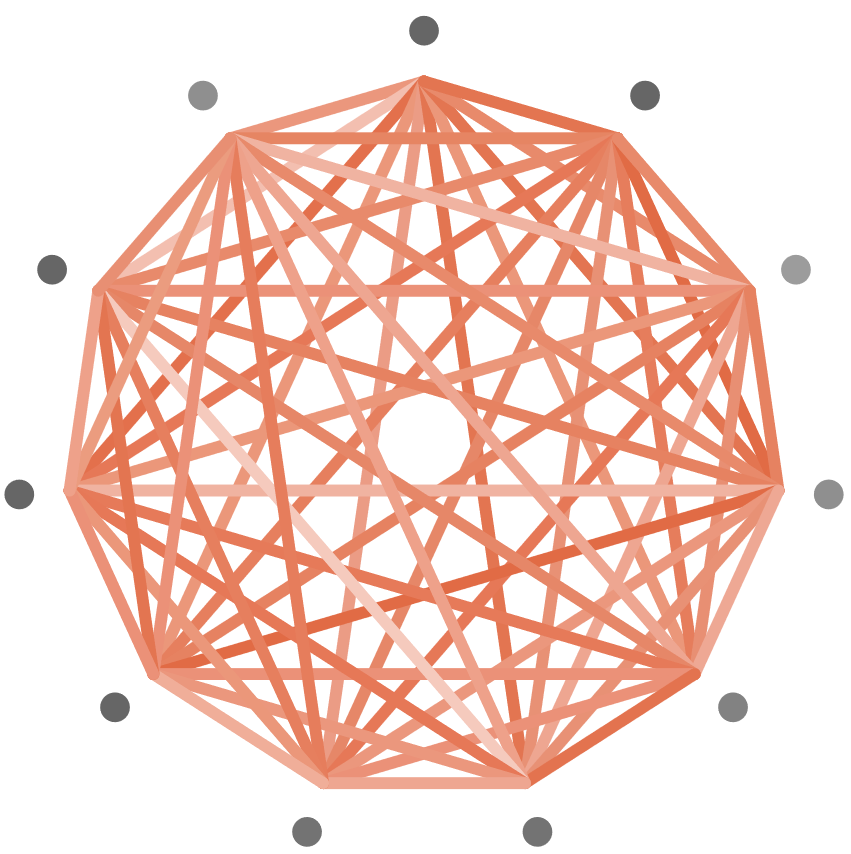}} \\
  \end{tabular} 
  \caption{The qubit topology and connectivities of (a) IBM Kolkata: A 27-qubit superconducting qubit quantum computer (b) Google Sycamore: A 53-qubit superconducting qubit quantum computer (c) An 11-qubit IonQ trapped-ion quantum computer. Note that the IonQ device has full connectivity of all the qubits in its system, but some of these connectivities have high noise levels.}
  \label{fig:qubit-topology}
\end{center}
\end{figure*}

The qubit topology graph of the device represents the qubit layout of the hardware. It shows how the physical qubits are connected on the real device. The qubit topology for some devices are shown in Figure \ref{fig:qubit-topology} \cite{ibm-quantum-experience}. This graph is important when a compiler tries to map a circuit to the quantum device, because it shows how qubits are connected and this is something that the compiler has to consider when transforming circuits to run in devices. For example, if two qubits are not physically connected, a CNOT gate cannot be executed between the two qubits. It is not possible to run any circuit on a real device without rewriting the circuit to match the device topology of the intended device to run on. This is part of a process called transpilation. Hence, topology graphs play an important role when designing circuits for algorithms in the NISQ-era. The qubits on IBM hardware are fixed. IBM uses superconducting transmon qubits rather than trapped-ion qubits. They don't move around as trapped-ion qubit hardware. Therefore, IBM Q quantum devices have limited connectivity among the qubits of the device. This property gives rise to the optimal qubit mapping problem as elaborated in the subsequent sections.

Due to noise, the depth and width of quantum circuits are limited. The low depth and width of quantum circuits mean that the capabilities of today's NISQ machines are dependent on the realization of hybrid algorithms. Hybrid algorithms contain both quantum and classical computing parts. At present all quantum algorithms follows this generic hybrid structure. The quantum part is like any quantum algorithm: the initial state is prepared, followed by unitary quantum gate operations, followed by measurements of the final states. The final state is measured by varying the input parameters of the quantum gate operations through classical post-processing till the results improve to the desired level.

Larger devices have significantly more noise than smaller ones. Hence, cutting a large circuit into smaller subcircuits seems to be a way to realize better fidelity. Many seemingly intractable problems for current NISQ-era devices are made possible by dividing the humongous problem-specific circuit into manageable ones that the current architectures can run. This improves \cite{saleem2021quantum} on the design of distributed quantum systems.

In a distributed quantum computing, multiple smaller non-local quantum devices are connected coherently, with all the nodes together working as a single large ecosystem of quantum computation.  The number of qubits scales linearly with the number of interconnected devices.  Hence, distributed Quantum Computing can act as an intermediate solution to the scaling-up problem. 

\section{Compiler Overview}

In the previous sections, we have observed that the NISQ-era devices are ones with low qubit count and high inherent device noise. The depth of a circuit is the number of time-steps required,  assuming that gates acting on distinct bits can operate simultaneously (that is, the depth is the maximum length of a directed path from the input to the output of the circuit) \cite{preskill_2015}. For NISQ-era devices, as the depth of the quantum circuit increases, the cumulative noise in the circuit also increases. The primary objective of our proposed compiler architecture is to reduce the depth of the quantum circuit by employing a recursive circuit cutting algorithm. This technique drastically reduces the size of the circuit to be executed. This, however, comes at a cost: The circuit reconstruction done as of now employs random shot sampling, thereby inducing finite sampling noise in the circuits. \textbf{Qurzon} tries to mitigate this noise by employing an optimal cut level algorithm, which finds the optimal level of recursive circuit cuts where the finite sampling noise is minimized, but maximizing the number of fragmented circuits, for ease of execution.

\begin{figure}[h!]
    \includegraphics[width=\linewidth]{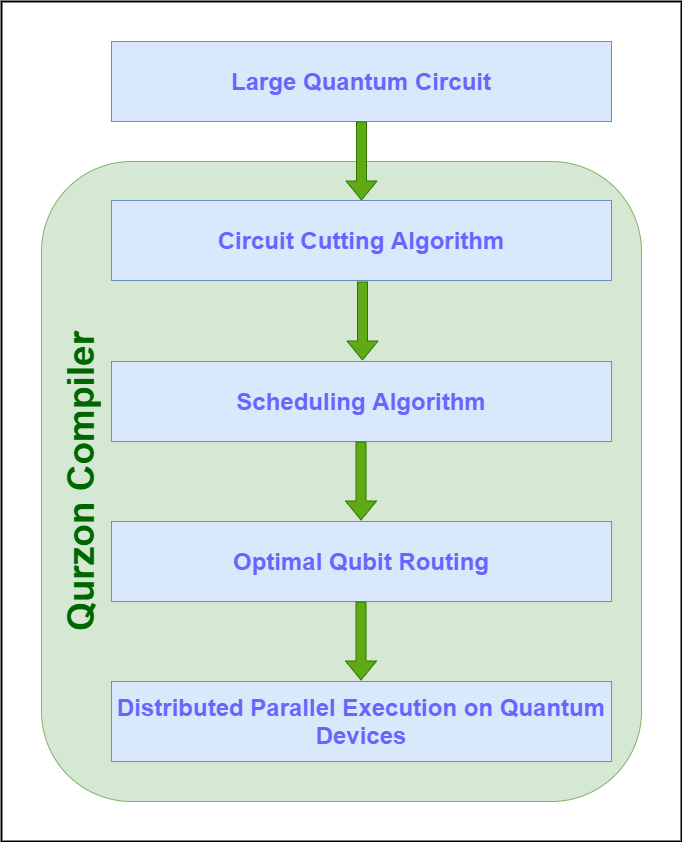}
    \caption{A high level overview of the \textbf{Qurzon} compiler, showing the different components and workflow}
    \label{fig:compiler-overview}
\end{figure}

A diagrammatic overview of \textbf{Qurzon}'s components can be found in Figure \ref{fig:compiler-overview}, which shall be briefly introduced here and elaborated on later in the subsequent subsections. 

\begin{enumerate}
    \item \textbf{Circuit Cutting Algorithm: } One of the primary components of the \textbf{Qurzon} is the circuit cutting algorithm. This takes a large multiqubit (>2) circuit and uses the CutQC \cite{tang2021cutqc} algorithm to find the optimal cut. The optimal cut finding is done using the Gurobi Solver with a mixed-integer programming formulation of the circuit's directed acyclic graph (DAG). This circuit cutting is employed recursively in order to break the circuit into multiple parts. Once done, it is passed on to the next part of the compiler.
   \item  \textbf{Scheduling Algorithm: } This part of the algorithm takes the cut circuits from the previous step and arranges them in a queue. It then takes into account the available quantum devices in the device pool. Once done, it allocates the subcircuits as jobs to the available quantum devices based on a greedy algorithm and passes it down to the next stage of the compilation process.
   \item  \textbf{Optimal Qubit Routing: } Once the scheduling algorithm zeros in on a target quantum device to run the subcircuit, the optimal qubit routing algorithm takes into account the current device qubit topology. Once done, it optimally places the quantum gates associated with the logical qubits of the subcircuit onto the physical qubits of the device, minimizing the number of SWAP gates required in the process.
  \item \textbf{Parallel Execution on Distributed Quantum Devices}: The subcircuits placed optimally on the quantum devices are then executed in parallel from the quantum device pool. As of now the device pool is considered to be a homogenous collection of superconducting qubit devices. 
  \item \textbf{Subsequent Reconstruction}: The entire circuit is consequently reconstructed using a shot-based probabilistic sampling of conditional probabilities of the subcircuit executions. This gives us the final output of the final quantum circuit.
\end{enumerate}
%

\subsection{Circuit Cutter}
The theory of circuit cutting has been first proposed in the paper by Peng et.al. \cite{peng2020simulating}. This section will be containing the mathematical framework for the circuit cutting algorithm, as proposed by Ayral et. al. \cite{ayral2020quantum} and thereafter, will be elaborating on the CutQC Algorithm \cite{tang2021cutqc} that \textbf{Qurzon} uses. 

\subsection*{Circuit Bipartitioning}

The calculation is a summary of the calculation presented in \cite{ayral2020quantum}. For further elaboration on this paper, the authors suggest referring to the original paper.

An $m$ qubit quantum circuit can be expressed as a composition of superoperators as follows: 

$$\mathcal{C} = \mathcal{C}^{\, after}_A \circ  \mathcal{C}^{\, after}_B \circ  \mathcal{C}^{\, before}_A \circ  \mathcal{C}^{\, before}_B$$

Here, the support of the super operators $\mathcal{C}^{\, after}_A$ and $\mathcal{C}^{\, after}_B$ is a bipartition of the quantum circuit. Similarly, the support of the superoperators $\mathcal{C}^{\, before}_A$ and $\mathcal{C}^{\, before}_B$ also forms a bipartition of the circuit $i.e.,$ without loss of generality, one can assume that the support of $\mathcal{C}^{\, before}_A$ refers to the qubits $q_0, q_1, \hdots,  q_{m-1}$ and that of $\mathcal{C}^{\, before}_B$ refers to $q_m, q_{m+1}, \hdots q_n$.

The final state of the circuit is given by the density matrix: $$\rho = \mathcal{C}(\rho)=\mathcal{C}^{\, after}_A \circ  \mathcal{C}^{\, after}_B \circ  \mathcal{C}^{\, before}_A \circ  \mathcal{C}^{\, before}_B(\rho_0)$$ where $\rho_0$ is the density matrix of the initial state. The  probability  of  measuring  a  state  $i$  with  binary  representation $i = (\hat{b}_{0}, \hat{b}_{1}, \hdots \hat{b}_{n-1})$ is given by: $$p(i) = Tr [ \Pi_i \cdot \rho ]$$ where $\Pi_i$ is the projector on the state $i$ $(i=0,\hdots, 2^m)$. 

Therefore,  $$p(i) = Tr [ \Pi^\dagger_i \cdot \mathcal{C}(\rho)=\mathcal{C}^{\, after}_A \circ  \mathcal{C}^{\, after}_B \circ \mathcal{C}^{\, before}_A \circ \mathcal{C}^{\, before}_B(\rho_0)]$$ as $\Pi^\dagger_i = \Pi_i$. Now if we switch to the Pauli basis, this results in the expression: $$p(i) = 2^m \langle \langle \Pi_i \, | \, \mathcal{R}^{\, after}_A \circ \mathcal{R}^{\, after}_B \circ \mathcal{R}^{\, before}_A \circ \mathcal{R}^{\, before}_B | \, \rho_0 \rangle \rangle$$ where $\mathcal{R}^{\, after/before}_{A/B}$ are the Pauli Transfer Matrices (PTM) representation of the superoperators as given by: $$[\mathcal{R}_{\alpha \beta}]  = \frac{1}{d}Tr[P_\alpha \cdot \mathcal{C}(P_\beta) ]$$ where $P_\alpha$ is a generalised Pauli matrix acting on $n$ qubits.

Now, the PTM representation of the Identity superoperator as follows: $$\mathcal{R_I} = \sum_{\alpha = X,Y,Z}\sum_{bb' = \lbrace 0,1 \rbrace} \widetilde{\gamma}^{bb'}_\alpha | \sigma_\alpha^b \rangle \rangle \langle \langle \sigma_\alpha^{b'} |,$$

where $| \sigma_\alpha^b \rangle \rangle$ are the (real) coordinates in the Pauli basis of the density matrix corresponding  to  the  $b^{th}$  eigenvector $| \psi_\alpha^b \rangle$ of Pauli matrix $\sigma_\alpha$. The $\widetilde{\gamma}$  tensor  is  given  by $\widetilde{\gamma}^{bb'}_X = \widetilde{\gamma}^{bb'}_Y = 2\delta_{bb'} - 1$ and $\widetilde{\gamma}^{bb'}_Z = 2\delta_{bb'}$.

On inserting $\mathcal{R_I}$ acting on the $n^{th}$ qubit $q_n$ and calculating the probability: 

\begin{multline*}
    p(i) = 2^m \langle \langle \Pi_i \, | \underbrace{\mathcal{R}^{\, after}_A}_{q_0, \hdots, q_{n-1}} \circ  \underbrace{\mathcal{R}^{\, after}_B}_{q_{n+1}, \hdots, q_{m-1}} \circ   \underbrace{\mathcal{R_I}}_{q_n} \\ 
    \circ  \underbrace{\mathcal{R}^{\, before}_A}_{q_0, \hdots, q_{n-1}} \circ  \underbrace{\mathcal{R}^{\, before}_B}_{q_{n+1} \hdots q_{m-1}} | \, \rho_0 \rangle \rangle
\end{multline*}

On applying the decomposition of $\mathcal{R_I}$ as given above: 
\begin{multline*}
    p(i) = \mathcal{R_I} = \sum_{\alpha = X,Y,Z}\sum_{bb' = \lbrace 0,1 \rbrace} \widetilde{\gamma}^{bb'}_\alpha \\
    \times \langle \langle \Pi_i |_{ \, q_0, \hdots , q_{n-1}} \langle \langle \Pi_i |_{ \, q_{n+1}, \hdots , q_{m-1}} \mathcal{R}^{\, after}_A \mathcal{R}^{\, after}_B | \, \sigma_\alpha^b \rangle \rangle_{\, {q_n}}\\
    \times \langle \langle \sigma_\alpha^{b'} |_{\, q_n} \mathcal{R}^{\, before}_A \mathcal{R}^{\, before}_B | \, \rho_0 \rangle \rangle_{\, q_0, \hdots ,{q_n}} | \, \rho_0 \rangle \rangle_{q_{n+1}, \hdots q_{m-1}}
\end{multline*}

Therefore, the final expression for the probabilities is given by: 

\begin{multline}
 p(\hat{b}_{0}\dots \hat{b}_{m-1})= \frac{1}{2}\sum _{\alpha =X,Y,Z}\sum _{bb'\in \{0,1\}}\tilde{\gamma }_{\alpha }^{bb'}p_{A}^{\alpha }\\
 (\hat{b}_{0}\dots \hat{b}_{n-1};b')\nonumber \times p_{B}^{\alpha b}(\hat{b}_{n}\dots \hat{b}_{m-1})   
\end{multline}

where,
$
p_{A}^{\alpha }(\hat{b}_{0}\dots \hat{b}_{n-1};b')\equiv 2^{n+1}\langle \langle \varPi _{\hat{b}_{0}\dots \hat{b}_{n-1}}|\langle \langle \sigma _{\alpha }^{b'}|_{q_{n}}\mathcal {R}_{A}|\rho _{0}\rangle \rangle _{q_{0}\dots q_{n}}
$, and,

$p_{B}^{\alpha b}(\hat{b}_{n}\dots \hat{b}_{m-1})\equiv 2^{m-n}\langle \langle \varPi _{\hat{b}_{n}\dots \hat{b}_{m-1}}|\mathcal {R}_{B}|\sigma _{\alpha }^{b}\rangle \rangle _{q_{n}}|\rho _{0}\rangle \rangle _{q_{n+1}\dots q_{m-1}}$,

The circuit cutting algorithm automatically locates optimal positions to cut a large quantum circuit into smaller subcircuits. The cutting algorithm is processed classically. This technique makes it possible to simulate quantum algorithms for large problems, which would be otherwise impossible considering the current device restrictions with scalability.There are many advantages to circuit cutting. The solution state remains the same even as the number of cuts increases. It is also relevant to tensor network contraction methods.  


\subsection{Scheduling Algorithm}
For this work, the scheduling algorithm takes into account the list of available devices the subcircuits can be executed upon. The scheduling algorithm allots the subcircuits to different devices (depending on their size and availability) while minimizing the loss of computational power and loss of time by making sure that most of the devices are used. The only function of the scheduler algorithm is to find the optimal distribution of the subcircuits over the list of devices.  

The salient features of the algorithm proposed herein consist of the following: 
\begin{itemize}
    \item The subcircuits that are maintained in a priority queue, with a user-defined priority assigned to them before enqueueing.
    \item By default the subcircuits are to have a priority zero. 
    \item The priorities assigned to each subcircuits add to Qurzon's distributed nature as given in Section 3.4.
    \item Circuits with higher priority get executed earlier than the circuits with lower priority.
    \item The initial allocation of the circuits to the respective quantum devices are to be done based on a priority queue.
    \item The estimated time in the execution queue, post initial allocation is calculated and corresponding wait times are to be calculated.
    \item These wait times and their corresponding devices are arranged in increasing order of wait times.
    \item Any circuit that hasn't been allocated to a quantum device post initial allocation are to be allocate thereafter, using a round-robin algorithm, starting from the device with the least wait time.
\end{itemize}

  

A brief outline of the algorithm is given as follows:

\begin{algorithm}
\caption{Proposed Scheduling Algorithm}\label{alg:cap}
\begin{algorithmic}
\Require List of subcircuits $C_i(p_i)$ to be executed, $C\_list$, where $p_i$ is the user-defined priority assigned to each circuit, List of available devices, $D\_list$
\State $A\_list$ = []
\While{$A\_list \neq D\_list$}
\State Sort($C\_list$) in decreasing order of $p_i$
\State Assign $C_i$ to quantum device $D_i$
\State Append $D_i$ to $A\_list$
\State Remove $C_i$ from $C\_list$
\If{$C\_list$ is not empty}
    \State Get estimated times $\tau_i$ for all $D_i$
    \State Append $D_i(\tau_i)$ to list WD
    \State Sort WD in ascending order
    \State Assign remaining circuits to each device in WD using a round-robin algorithm
\EndIf
\EndWhile
\end{algorithmic}
\end{algorithm}

\begin{figure*}[h!]
\begin{center}
    \begin{tabular}{ccc}
        \subfloat[]{\includegraphics[width = 2.1in]{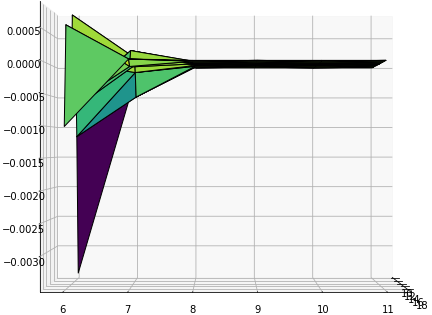}} &
        \subfloat[]{\includegraphics[width = 2.1in]{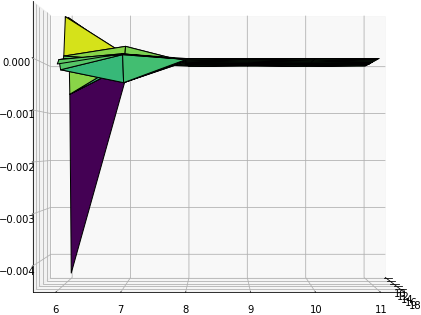}} &
        \subfloat[]{\includegraphics[width = 2.1in]{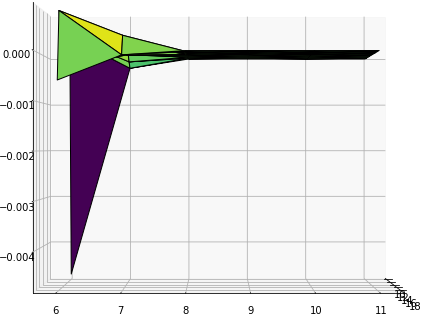}}
  \end{tabular}
  \begin{tabular}{ccc}
    \subfloat[]{\includegraphics[width=2.1in]{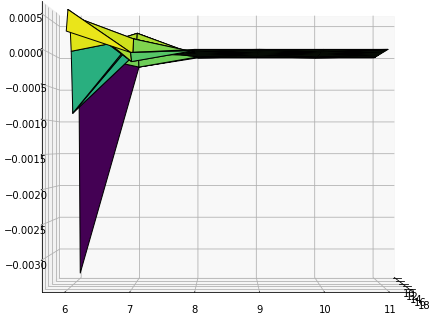}} &
    \subfloat[]{\includegraphics[width=2.1in]{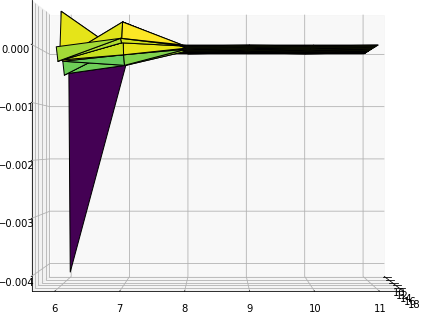}} &
    \subfloat[]{\includegraphics[width=2.1in]{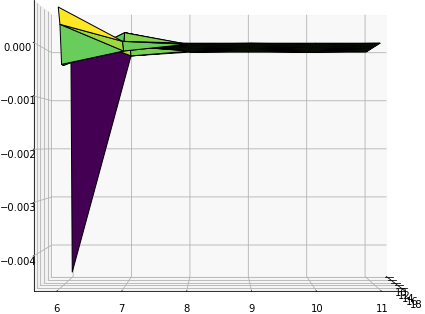}}
  \end{tabular}
  \caption{Cross-sectional view of the figures plotted from the results. The y-axis represents the difference $E - E'$, where $E$ is the mean squared error of the circuit without $t|ket\rangle$ and $E'$ is the mean squared error of the same using $t|ket\rangle$.The x-axis indicates the size of the input circuit. The circuits used were (a) Grover's algorithm (b) Approximated quantum fourier transform (AQFT) (c) Quantum Fourier Transform (QFT) (d) Supremacy Circuit (e) Linear Supremacy Circuit (f) Random Circuit}
  \label{fig:Cross-sectional}
\end{center}
\end{figure*}

\subsection{Optimal Qubit Routing}
As introduced in section 2, NISQ-era devices are inherently susceptible to noise. NISQ devices also have limited qubit connectivity; this is particularly prevalent not only in superconducting qubit architectures but also in trapped-ion systems, as shown in Figure \ref{fig:qubit-topology}. Although the qubits in the trapped-ion systems appear to be fully connected, something to keep in mind is that in such architectures, the connectivity between the qubits is also prone to noise. This translates to the fact that not all connections have the same fidelity. This limited connectivity in devices gives birth to the optimal qubit routing problem. 

The qubit routing problem is fundamental for circuit execution in NISQ-era devices. The motivation for the problem is to map a logical qubit, a qubit allocated in the circuit or the algorithm must translate directly to a physical qubit, a qubit on the actual quantum device. In doing so, the quantum circuit must go through transpilation or a source-to-source translater. What this stage does is, taking a quantum circuit into account, the transpiler decomposes the same into an equivalent circuit with a set of gates that can be run on the quantum device. Something to note is that the set of gates or the gate set is different for every quantum device. 


Once the transpiler does its job, a set of gates is left behind to be executed on the device. These include single-qubit as well as two-qubit gates. The single-qubit gates aren't much of a concern since they can be mapped to the individual physical qubits with ease. However, to execute the two-qubit gates, the qubit connectivity of the devices are accounted for. This means that any two logical qubits that are adjacent may not be physically adjacent to each other. 

A viable solution to the problem of having non-adjacent interacting qubits is to insert SWAP gates to exchange it with a neighbouring qubit, thereby moving it closer to the desired qubit position. As this method introduces new gates into the quantum circuit, the depth of the circuit increases thereby compromising the algorithm's performance. This problem is called the qubit routing problem, wherein two desired qubits are made to interact with one another by minimizing the number of SWAP gates necessary to do so. This problem is shown to be equivalent to the token-swapping problem and is proven to be at least NP-hard and possibly PSPACE-complete \cite{siraichi2018qubit}. 

However, efforts have been made to design an exact solution to the qubit routing problem. Siraichi et al. \cite{siraichi2018qubit} proposed a dynamic programming method, resulting in a complexity that is exponential to the number of qubits, and a heuristic solution that approximates solutions to small circuits reasonably well. Zulehner et al. \cite{zulehner2018efficient} proposed an algorithm that takes into account partitions based on the depth of the circuit and using the A* algorithm to search for optimal solutions, specializing in IBM devices. There is considerable literature towards finding an optimal solution to this problem \cite{wille2019mapping} \cite{nash2020quantum} \cite{tan2020optimal} , but here the approach proposed by $t|ket\rangle$ has been incorporated in \textbf{Qurzon} as outlined below: 

\subsubsection*{Optimal Layout Synthesis with $t|ket\rangle$}
$t|ket\rangle$ \cite{cowtan2019qubit} \cite{cowtan2020generic} \cite{Cowtan_2020} is a language-agnostic optimising compiler package that manages executable circuits and optimizes for physical qubit layout while also reducing the number of required operations. The system consists of two main components: a powerful optimizing compiler written in C++, and a lightweight user interface and runtime system written in Python. The python Interface is used to interact with a wide range of hardware, handle the full cycle of compilation, dispatch and results retrieval. The compilation steps can be controlled via the interface to suit the nature of the quantum circuit being run.  $t|ket\rangle$ deals with the qubit routing problem by using a heuristic method that produces optimal mapping systems in terms of depth, total gate count, and low run times. Circuit optimisation is especially pertinent on so-called noisy intermediate-scale quantum (NISQ) devices.  The longer the computation runs, the more noise builds up. It is specifically designed for NISQ devices, and includes features that minimise the influence of device errors on computation. $t|ket\rangle$ routing algorithm ensures the compilation of any quantum circuit to any existing device as long as its architecture can be represented as a simple connected graph. The algorithm works in four parts: decomposing the input circuit into timesteps; initial mapping; routing across timesteps; and swap synthesis and final rewriting based on device architecture constraints. 

\subsection{Parallel Execution on Distributed Quantum Systems}
As outlined in the scheduling algorithm, Qurzon is to be equipped with the power of distributed quantum computing, with the subcircuits being executed in parallel. The quantum devices used herein are assumed to be superconducting qubit devices, taking into account their respective decoherence noise, device noise, readout error rates, etc. With the ability to assign a user-defined priority to each subcircuit, Qurzon offers some degree of control over which circuits are assigned to which quantum device. In future iterations of this research, the authors of this paper aim to provide a robust abstraction to assign each subcircuit manually to each quantum device. Furthermore, the authors of this paper aim to extend the cross-platform functionality of this compiler to trapped-ion and photonic systems, thereby allowing researchers to execute different parts of a quantum algorithm on different device architectures. The authors predict that the significant challenges faced herein are calibrating the measurement readouts to construct a full algorithm. The authors hope to address this problem through subsequent research in the field.

\subsection{Stitching Quantum Circuits Post Execution}
Circuit reconstruction/stitching is a classical post processing method that can reconstruct the output of the original circuit. It reconstructs the probability distribution over measurement outcomes for the original quantum circuit from probability distributions associated with the subcircuits. The circuit reconstruction process has a relatively low computational cost that makes the whole algorithm a practical alternative strategy for large quantum simulations. 

However that being said, the reconstruction algorithm used herein uses CutQC's \cite{tang2021cutqc} Dynamic Definition Query which still uses exponential time and resource complexity. This leaves room for improvement in subsequent iterations of this research wherein, attempts shall be made to reduce it to sub-exponential costs.

\section{Experimental Results}

\subsection{Experimental Setup}
The workflow for the "control system" $i.e.,$ running the circuits without using optimal layout synthesis is as follows: The circuits were generated using the code for the helper function in Wei Tang's CutQC paper \cite{tang2021cutqc}. Mock backends were taken from the IBM Qiskit Aer simulator library \cite{Qiskit}. The backends used were  Fake Tokyo, Fake Tenerife, Fake Melbourne, Fake Vigo, Fake Rueschlikon and Fake Poughkeepsie. Noise models for the respective backends were used and subjected to the scheduling algorithm. The circuit cutter used Gurobi's mixed integer programming solver and used 8192 shots for the reconstruction phase. Once the cut circuits have been run on the respective backends, they are reconstructed and their fidelities are noted. For the "experiment workflow" all the steps in the previous paragraph were done except when using the backends. Herein, the $t|ket\rangle$ backends were used from $t|ket\rangle$ IBM extension package. Here, $t|ket\rangle$ implements optimal layout synthesis while running the cut circuits and uses CutQC's reconstruction methods and generates the respective fidelities. 

\subsection{Case Study: The Bernstein-Vazirani Algorithm}
The Bernstein Vazirani algorithm was first introduced in \cite{doi:10.1137/S0097539796300921}. This algorithm is an extension to the Deutsch Josza algorithm and demonstrated the use of quantum computers to solve more problems more complex in nature than that of the Deutsch Josza. Here, we are given a oracle that outputs a 0 or a 1  i.e. $f(\{x_0,x_1,x_2,...\}) \rightarrow 0 \textrm{ or } 1 \textrm{ where } x_i \textrm{ is }0 \textrm{ or } 1$ along with a input string $s$. That means the oracle looks like $f(x) = s \cdot x \, \text{(mod 2)}$. The oracle returns with certainty with just one run of the function. The Bernstein Vazirani algorithm is used to find the secret input string $s$.
\\\\
For an elementary application of the Bernstein Vazirani algorithm, the circuit is given by:

\scalebox{0.75}{
\Qcircuit @C=1.0em @R=0.2em @!R { \\
	 	\nghost{ {q}_{0} :  } & \lstick{ {q}_{0} :  } & \gate{\mathrm{H}} & \qw & \ctrl{9} & \gate{\mathrm{H}} & \qw & \qw & \qw & \qw & \qw & \qw & \qw & \qw & \qw & \qw\\ 
	 	\nghost{ {q}_{1} :  } & \lstick{ {q}_{1} :  } & \gate{\mathrm{H}} & \qw & \qw & \ctrl{8} & \gate{\mathrm{H}} & \qw & \qw & \qw & \qw & \qw & \qw & \qw & \qw & \qw\\ 
	 	\nghost{ {q}_{2} :  } & \lstick{ {q}_{2} :  } & \gate{\mathrm{H}} & \qw & \qw & \qw & \ctrl{7} & \gate{\mathrm{H}} & \qw & \qw & \qw & \qw & \qw & \qw & \qw & \qw\\ 
	 	\nghost{ {q}_{3} :  } & \lstick{ {q}_{3} :  } & \gate{\mathrm{H}} & \qw & \qw & \qw & \qw & \ctrl{6} & \gate{\mathrm{H}} & \qw & \qw & \qw & \qw & \qw & \qw & \qw\\ 
	 	\nghost{ {q}_{4} :  } & \lstick{ {q}_{4} :  } & \gate{\mathrm{H}} & \qw & \qw & \qw & \qw & \qw & \ctrl{5} & \gate{\mathrm{H}} & \qw & \qw & \qw & \qw & \qw & \qw\\ 
	 	\nghost{ {q}_{5} :  } & \lstick{ {q}_{5} :  } & \gate{\mathrm{H}} & \qw & \qw & \qw & \qw & \qw & \qw & \ctrl{4} & \gate{\mathrm{H}} & \qw & \qw & \qw & \qw & \qw\\ 
	 	\nghost{ {q}_{6} :  } & \lstick{ {q}_{6} :  } & \gate{\mathrm{H}} & \qw & \qw & \qw & \qw & \qw & \qw & \qw & \ctrl{3} & \gate{\mathrm{H}} & \qw & \qw & \qw & \qw\\ 
	 	\nghost{ {q}_{7} :  } & \lstick{ {q}_{7} :  } & \gate{\mathrm{H}} & \qw & \qw & \qw & \qw & \qw & \qw & \qw & \qw & \ctrl{2} & \gate{\mathrm{H}} & \qw & \qw & \qw\\ 
	 	\nghost{ {q}_{8} :  } & \lstick{ {q}_{8} :  } & \gate{\mathrm{H}} & \qw & \qw & \qw & \qw & \qw & \qw & \qw & \qw & \qw & \ctrl{1} & \gate{\mathrm{H}} & \qw & \qw\\ 
	 	\nghost{ {q}_{9} :  } & \lstick{ {q}_{9} :  } & \gate{\mathrm{X}} & \gate{\mathrm{H}} & \targ & \targ & \targ & \targ & \targ & \targ & \targ & \targ & \targ & \gate{\mathrm{H}} & \qw & \qw\\ 
\\ }}

Demonstrated herein is the present state of the compiler and its inner workings traced as it passes through each step in the compiler. 

\subsubsection*{Circuit Cutting Phase}
Using Mixed Integer Programming model this circuit is optimally subdivided into several subcircuits which are given below. 

These subcircuits are then into scheduler algorithm which determines the architecture of the devices on which the subcircuits are to be run.

\scalebox{1.5}{
\Qcircuit @C=1.0em @R=0.2em @!R { \\
	 	\nghost{ {q}_{0} :  } & \lstick{ {q}_{0} :  } & \gate{\mathrm{H}} & \qw & \ctrl{2} & \gate{\mathrm{H}} & \qw & \qw & \qw\\ 
	 	\nghost{ {q}_{1} :  } & \lstick{ {q}_{1} :  } & \gate{\mathrm{H}} & \qw & \qw & \ctrl{1} & \gate{\mathrm{H}} & \qw & \qw\\ 
	 	\nghost{ {q}_{2} :  } & \lstick{ {q}_{2} :  } & \gate{\mathrm{X}} & \gate{\mathrm{H}} & \targ & \targ & \qw & \qw & \qw\\ 
\\ }}

\scalebox{1.5}{
\Qcircuit @C=1.0em @R=0.2em @!R { \\
	 	\nghost{ {q}_{0} :  } & \lstick{ {q}_{0} :  } & \gate{\mathrm{H}} & \ctrl{1} & \gate{\mathrm{H}} & \qw & \qw\\ 
	 	\nghost{ {q}_{1} :  } & \lstick{ {q}_{1} :  } & \qw & \targ & \gate{\mathrm{H}} & \qw & \qw\\ 
\\ }}

\scalebox{1.5}{
\Qcircuit @C=1.0em @R=0.2em @!R { \\
	 	\nghost{ {q}_{0} :  } & \lstick{ {q}_{0} :  } & \gate{\mathrm{H}} & \ctrl{2} & \gate{\mathrm{H}} & \qw & \qw & \qw\\ 
	 	\nghost{ {q}_{1} :  } & \lstick{ {q}_{1} :  } & \gate{\mathrm{H}} & \qw & \ctrl{1} & \gate{\mathrm{H}} & \qw & \qw\\ 
	 	\nghost{ {q}_{2} :  } & \lstick{ {q}_{2} :  } & \qw & \targ & \targ & \qw & \qw & \qw\\ 
\\ }}

\scalebox{1.5}{
\Qcircuit @C=1.0em @R=0.2em @!R { \\
	 	\nghost{ {q}_{0} :  } & \lstick{ {q}_{0} :  } & \gate{\mathrm{H}} & \ctrl{2} & \gate{\mathrm{H}} & \qw & \qw & \qw\\ 
	 	\nghost{ {q}_{1} :  } & \lstick{ {q}_{1} :  } & \gate{\mathrm{H}} & \qw & \ctrl{1} & \gate{\mathrm{H}} & \qw & \qw\\ 
	 	\nghost{ {q}_{2} :  } & \lstick{ {q}_{2} :  } & \qw & \targ & \targ & \qw & \qw & \qw\\ 
\\ }}

\scalebox{1.5}{
\Qcircuit @C=1.0em @R=0.2em @!R { \\
	 	\nghost{ {q}_{0} :  } & \lstick{ {q}_{0} :  } & \gate{\mathrm{H}} & \ctrl{2} & \gate{\mathrm{H}} & \qw & \qw & \qw\\ 
	 	\nghost{ {q}_{1} :  } & \lstick{ {q}_{1} :  } & \gate{\mathrm{H}} & \qw & \ctrl{1} & \gate{\mathrm{H}} & \qw & \qw\\ 
	 	\nghost{ {q}_{2} :  } & \lstick{ {q}_{2} :  } & \qw & \targ & \targ & \qw & \qw & \qw\\ 
\\ }}

\subsubsection*{Scheduling Phase}
During the scheduling phase of the compiler, the authors have used a naive scheduling algorithm which allows for the execution of the quantum circuit on one of the devices in the device pool. The naive scheduler used herein has scheduled the subcircuits to be executed on the IBM FakeTenerife. The qubit topology of the device is given as follows: 

\begin{figure}[h!]
    \includegraphics[width=\linewidth]{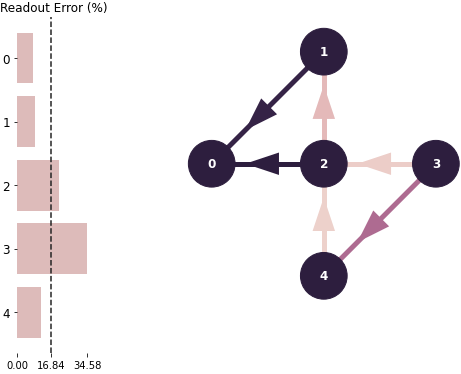}
    \caption{Qubit Topology of the IBM FakeTenerife mock backend}
    \label{fig:FT}
\end{figure}

However, in future Qurzon shall be leveraging the scheduling algorithm proposed herein. 

\subsection*{Qubit Mapping and Optimal Layout Synthesis}
Using the qubit topology of IBM FakeTenerife as shown in figure \ref{fig:FT}, the optimal layout synthesis was carried out using $t|ket\rangle$. $t|ket\rangle$ takes into account the connectivities between the qubits and have accordingly placed the quantum gates such that the execution is optimal and the number of SWAP gates required is minimized. 
\scalebox{1.4}{
\Qcircuit @C=1.0em @R=0.2em @!R { \\
	 	\nghost{ {q}_{0} :  } & \lstick{ {q}_{0} :  } & \gate{\mathrm{H}} & \qw & \qw & \ctrl{1} & \gate{\mathrm{H}} & \qw & \qw\\ 
	 	\nghost{ {q}_{1} :  } & \lstick{ {q}_{1} :  } & \gate{\mathrm{X}} & \gate{\mathrm{H}} & \targ & \targ & \qw & \qw & \qw\\ 
	 	\nghost{ {q}_{2} :  } & \lstick{ {q}_{2} :  } & \gate{\mathrm{H}} & \qw & \ctrl{-1} & \gate{\mathrm{H}} & \qw & \qw & \qw\\ 
\\ }}

\scalebox{1.5}{
\Qcircuit @C=1.0em @R=0.2em @!R { \\
	 	\nghost{ {q}_{0} :  } & \lstick{ {q}_{0} :  } & \qw & \targ & \gate{\mathrm{H}} & \qw & \qw\\ 
	 	\nghost{ {q}_{1} :  } & \lstick{ {q}_{1} :  } & \gate{\mathrm{H}} & \ctrl{-1} & \gate{\mathrm{H}} & \qw & \qw\\ 
\\ }}

\scalebox{1.4}{
\Qcircuit @C=1.0em @R=0.2em @!R { \\
	 	\nghost{ {q}_{0} :  } & \lstick{ {q}_{0} :  } & \gate{\mathrm{H}} & \qw & \ctrl{1} & \gate{\mathrm{H}} & \qw & \qw\\ 
	 	\nghost{ {q}_{1} :  } & \lstick{ {q}_{1} :  } & \qw & \targ & \targ & \qw & \qw & \qw\\ 
	 	\nghost{ {q}_{2} :  } & \lstick{ {q}_{2} :  } & \gate{\mathrm{H}} & \ctrl{-1} & \gate{\mathrm{H}} & \qw & \qw & \qw\\ 
\\ }}

\scalebox{1.4}{
\Qcircuit @C=1.0em @R=0.2em @!R { \\
	 	\nghost{ {q}_{0} :  } & \lstick{ {q}_{0} :  } & \gate{\mathrm{H}} & \qw & \ctrl{1} & \gate{\mathrm{H}} & \qw & \qw\\ 
	 	\nghost{ {q}_{1} :  } & \lstick{ {q}_{1} :  } & \qw & \targ & \targ & \qw & \qw & \qw\\ 
	 	\nghost{ {q}_{2} :  } & \lstick{ {q}_{2} :  } & \gate{\mathrm{H}} & \ctrl{-1} & \gate{\mathrm{H}} & \qw & \qw & \qw\\ 
\\ }}

\scalebox{1.4}{
\Qcircuit @C=1.0em @R=0.2em @!R { \\
	 	\nghost{ {q}_{0} :  } & \lstick{ {q}_{0} :  } & \gate{\mathrm{H}} & \qw & \ctrl{1} & \gate{\mathrm{H}} & \qw & \qw\\ 
	 	\nghost{ {q}_{1} :  } & \lstick{ {q}_{1} :  } & \qw & \targ & \targ & \qw & \qw & \qw\\ 
	 	\nghost{ {q}_{2} :  } & \lstick{ {q}_{2} :  } & \gate{\mathrm{H}} & \ctrl{-1} & \gate{\mathrm{H}} & \qw & \qw & \qw\\ 
\\ }}

\subsubsection*{Circuit Reconstruction Phase}
The subcircuits thus formed and executed on quantum device are consequently reconstructed using Dynamic Definition Query as outlined in \cite{tang2021cutqc}.

The measurement thus done used 1024 shots to reconstruct the probability landscape of the entire quantum circuit. This gives rise to finite sampling error which can be quantified in the results using the following parameters and their corresponding values as outlined in the table below. 

\begin{table}[h!]
\centering
\begin{tabular}{|l|l|} 
\hline
\textbf{Metric }                        & \textbf{Value}        \\ 
\hline
Cross Entropy                  & 2.402524222  \\ 
\hline
Mean Absolute Percentage Error & 8.88194E+14  \\ 
\hline
Mean Squared Error             & 0.000815706  \\ 
\hline
$\chi^2$ Error                 & 1.668078342  \\ 
\hline
Heavy output probability       & 0.09048925   \\
\hline
\end{tabular}
\hfill \break
\caption{\label{tab:table-name}Table showing different errors for the \\
10 Qubit Bernstein-Vazirani Algorithm}
\end{table}





\subsection{Results and Discussions}
The results published herein are based on a limited number of qubits, with limited depth, due to computational power constraints. However, keeping in mind, the prototypical nature of this project, the results are sufficient to draw meaningful inference, paving the way for future work in the field.

\begin{figure}[h!]
    \includegraphics[width=\linewidth]{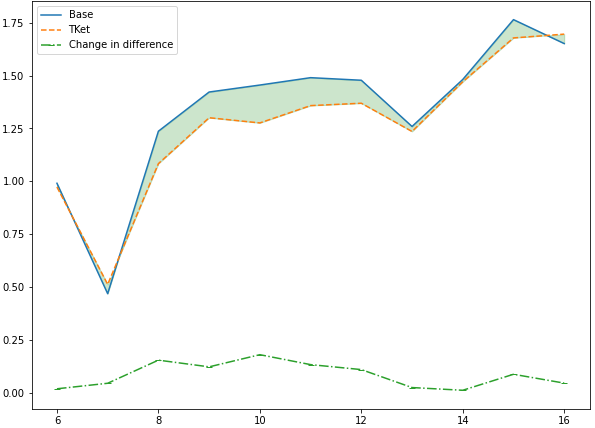}
    \caption{A comparison of the $\chi^2$ error of the CutQC circuits with and without $t|ket\rangle$ for the Bernstein Vazirani algorithm}
    \label{fig:compBV}
\end{figure}

On benchmarking several circuits, namely the Bernstein Vazirani algorithm, Grover's algorithm, the AFQT, the QFT, the supremacy circuit, the linear supremacy circuit and random circuits, through the pipeline proposed in this paper, a general trend can be seen. The results of these benchmarks can be observed in Figure \ref{fig:Cross-sectional} and \ref{fig:compBV}. The y-axis consists of the difference in the fidelities $\delta_F = (E-E')$, where $E$ is the mean squared error of the circuit without subjecting it to a qubit mapping algorithm and $E'$ is the mean squared error of the circuit after $t|ket\rangle$ was applied. The x-axis consists of the number of qubits in the input circuits. 

When the size of the input circuits is small, the $\delta_F$ tends to stay negative. This is because, when circuit size is small and the size of the cut circuits are small, the number of cuts in the circuits are minimal.  The cut circuits are then optimally routed using the $t|ket\rangle$ module. On the control side of these experiments, the optimal layout synthesis module was not used. It was noted that at these levels, optimal qubit mapping plays a significant role in mitigating the finite sampling noise, found as a result of circuit reconstruction. At large qubit counts, with cut size remaining constant, the reconstruction noise is great enough to nullify the effects of the qubit mapping. 

This is an important observation because although due to computational constraints large circuits could not be simulated, a hypothesis can be reached, and can be explored further in future iterations of this research. The hypothesis is that when large circuits can be made to run through the compiler, with large enough cut circuit size (say, equal to the size of device register), qubit routing is to play a pivotal role in mitigating circuit reconstruction noise, thereby significantly improving the fidelity of the circuits formed after reconstruction. 

Furthermore, the depth of the quantum circuit plays minimal impact on the fidelity, with or without the qubit routing scheme. This is due to the depth-agnostic nature of the $t|ket\rangle$ algorithm. However depth-aware mapping techniques maybe used in future iterations of this research in order to study the depths therein. Furthermore, it remains to be seen if the depth can be reduced effectively by using techniques similar to CutQC \cite{tang2021cutqc}. Such tomography inspired techniques have been implemented by Perlin et. al. \cite{perlin2021quantum} but remains to be studied for general circuits. 

\subsection*{Running time of the circuits}
This section attempts to provide a formalism regarding the running times of the circuits on Qurzon, as it stands, in the latest iteration. This formalism is heavily dependent on the CutQC algorithm and the availability of quantum devices along with classical resources. 

For this analysis, $n$ quantum subcircuits, post circuit cutting, are to be considered to be allocated to run on $k$ quantum devices. Instead of queuing the quantum circuits, the authors herein propose running the circuits parallelly on the quantum devices, subject to device availability. Let $S_k$ denote the $k^{th}$ quantum processor, that is available to be used and $t(S_k)$ denote the time taken to run the circuit  on the $k^{th}$ quantum processor. Without loss of generality, it is assumed that $n>k$. The time taken for classical post processing is considered as $C(S_i)$ for the $i^{th}$ subcircuit. 

The time, therefore, taken for the classical post processing of the $n$ subcircuits is given by $T_c = \sum_{i=1}^n C(S_i)$. 

The time taken for the circuits to be executed on the quantum processor be given by $T_q = max( t(S_1) , t(S_2) , \dots , t(S_n)   )$. This is due to \textbf{Qurzon's} proposed scheduling algorithm that shall allow for the parallel execution of these subcircuits. 

Since $n>k$, it might be that some subcircuit might require some wait time before some quantum device is allocated. This wait time is characterised by $W(S_i)$ for the $i^{th}$ subcircuit in waiting. Note that $W(S_i)$ can be zero, for the circuits that can be associated with a quantum processor. Hence, the total wait time is given by $T_w = \sum_{i=1}^n W(S_i)$.

Therefore, the total time taken for the end to end execution of a quantum circuit, that has been divided into $n$ subcircuits associated with $k$ quantum devices is given by: 

$$T = T_q + T_c + T_w$$
\begin{multline*}
    T = max( t(S_1) , t(S_2) , \dots , t(S_n)   ) + \sum_{i=1}^n C(S_i) + \sum_{i=1}^n W(S_i)
\end{multline*}

\subsection*{Circuit cutting to reduce multiqubit gates}

Every quantum circuit consists of unitary operators that operate over either one qubit or multiple qubits. In most quantum devices, multi-qubit gates are decomposed into several two-qubit gates in order to be accommodated in the gate-set for a quantum device. The gate-set for a quantum device is defined to be the set of operations that a quantum processor is to physically run on itself, in order to output the results of a circuit. This process is called transpilation and is handled by a piece of software called the transpiler. At the hardware level, therefore, the fidelity of the circuit is dependent on the types of gates that the circuit uses. Specifically, at the juxtaposition of logic synthesis and execution, the two-qubit gates are responsible \cite{PhysRevLett.113.220502} \cite{PhysRevApplied.10.054062} \cite{PhysRevApplied.12.054023} \cite{PhysRevApplied.14.024070} \cite{PhysRevA.102.022619} for the fidelity of the quantum circuits. The two-qubit gates are also the motivation behind the optimal qubit mapping problem for NISQ-era devices. 

The beauty of Qurzon is that due to the circuit cutting techniques, the number of two-qubit gates in the circuit is highly diminished when the size of the qubit register of the quantum device is small. The tradeoff, however, lies in the fact that whenever the size of the qubit register for the cut circuits is low, the time taken for classical post-processing, exponentially increases. This has been shown by Tang et. al. in the CutQC paper \cite{tang2021cutqc}. Not only that, the reconstruction noise after execution for several of these minuscule circuits exceeds the noise induced by these two-qubit gates. The authors of this paper, therefore, feel, that it is imperative for the reconstruction methodology for these subcircuits to be improved upon such that the probability landscape of large circuits can be calculated effectively, swiftly and reliably. This point also remains to be worked on by the authors, in the successive iterations of this project.

\section{Conclusion and Future Scope}
This piece of work introduces \textbf{Qurzon}, a novel approach to compiling and executing circuits on NISQ-era devices. It also studies the effects of qubit routing on the quantum divide and compute approach. It is noted that for sufficiently large circuits, qubit routing plays a major role in determining the circuit's fidelity after being subjected to the cut and execute methodology. Furthermore, this work proposes the use of parallel scheduling algorithms for quantum multi-computing. The work proposed herein is to be extended to include devices from multiple vendors such as Google, IBM, Rigetti, Xanadu, IonQ i.e. on photonic and trapped ion hardware. The authors of this paper are currently investigating methods to cut down circuit depth that is to be integrated in future iterations of this work. Additionally this work paves the way for future research into areas wherein different parts of a quantum circuit maybe executed on different hardware and calibrated accordingly. Another major scope for future research in this field is to reduce the classical post-processing to a sub-exponential computational cost.



\normalsize
\bibliography{references}

\begin{thebibliography}{43}
\providecommand{\natexlab}[1]{#1}
\providecommand{\url}[1]{\texttt{#1}}
\expandafter\ifx\csname urlstyle\endcsname\relax
  \providecommand{\doi}[1]{doi: #1}\else
  \providecommand{\doi}{doi: \begingroup \urlstyle{rm}\Url}\fi

\bibitem[Nielsen and Chuang(2010)]{nielsen_chuang_2010}
Michael~A. Nielsen and Isaac~L. Chuang.
\newblock \emph{Quantum Computation and Quantum Information: 10th Anniversary
  Edition}.
\newblock Cambridge University Press, 2010.
\newblock \doi{10.1017/CBO9780511976667}.

\bibitem[Preskill(2018)]{preskill2018quantum}
John Preskill.
\newblock Quantum {C}omputing in the {NISQ} {E}ra and {B}eyond.
\newblock \emph{Quantum}, 2:\penalty0 79, 2018.

\bibitem[Steane(1998)]{steane1998quantum}
Andrew Steane.
\newblock Quantum computing.
\newblock \emph{Reports on Progress in Physics}, 61\penalty0 (2):\penalty0 117,
  1998.

\bibitem[Williams(2010)]{williams2010explorations}
Colin~P Williams.
\newblock \emph{Explorations in quantum computing}.
\newblock Springer Science \& Business Media, 2010.

\bibitem[Biamonte et~al.(2017)Biamonte, Wittek, Pancotti, Rebentrost, Wiebe,
  and Lloyd]{biamonte2017quantum}
Jacob Biamonte, Peter Wittek, Nicola Pancotti, Patrick Rebentrost, Nathan
  Wiebe, and Seth Lloyd.
\newblock Quantum machine learning.
\newblock \emph{Nature}, 549\penalty0 (7671):\penalty0 195--202, 2017.

\bibitem[Schuld et~al.(2015)Schuld, Sinayskiy, and
  Petruccione]{schuld2015introduction}
Maria Schuld, Ilya Sinayskiy, and Francesco Petruccione.
\newblock An introduction to quantum machine learning.
\newblock \emph{Contemporary Physics}, 56\penalty0 (2):\penalty0 172--185,
  2015.

\bibitem[Cao et~al.(2019)Cao, Romero, Olson, Degroote, Johnson, Kieferov{\'a},
  Kivlichan, Menke, Peropadre, Sawaya, et~al.]{cao2019quantum}
Yudong Cao, Jonathan Romero, Jonathan~P Olson, Matthias Degroote, Peter~D
  Johnson, M{\'a}ria Kieferov{\'a}, Ian~D Kivlichan, Tim Menke, Borja
  Peropadre, Nicolas~PD Sawaya, et~al.
\newblock Quantum chemistry in the age of quantum computing.
\newblock \emph{Chemical reviews}, 119\penalty0 (19):\penalty0 10856--10915,
  2019.

\bibitem[Dasgupta and Humble(2020)]{dasgupta2020characterizing}
Samudra Dasgupta and Travis~S Humble.
\newblock Characterizing the stability of nisq devices.
\newblock In \emph{2020 IEEE International Conference on Quantum Computing and
  Engineering (QCE)}, pages 419--429. IEEE, 2020.

\bibitem[Cowtan et~al.(2019)Cowtan, Dilkes, Duncan, Krajenbrink, Simmons, and
  Sivarajah]{cowtan2019qubit}
Alexander Cowtan, Silas Dilkes, Ross Duncan, Alexandre Krajenbrink, Will
  Simmons, and Seyon Sivarajah.
\newblock On the qubit routing problem.
\newblock \emph{arXiv preprint arXiv:1902.08091}, 2019.

\bibitem[Ferrari et~al.(2021)Ferrari, Cacciapuoti, Amoretti, and
  Caleffi]{Ferrari_2021}
Davide Ferrari, Angela~Sara Cacciapuoti, Michele Amoretti, and Marcello
  Caleffi.
\newblock Compiler design for distributed quantum computing.
\newblock \emph{IEEE Transactions on Quantum Engineering}, 2:\penalty0 1–20,
  2021.
\newblock ISSN 2689-1808.
\newblock \doi{10.1109/tqe.2021.3053921}.
\newblock URL \url{http://dx.doi.org/10.1109/TQE.2021.3053921}.

\bibitem[Ferrari et~al.()Ferrari, Nasturzio, and Amoretti]{ferrarisoftware}
D~Ferrari, S~Nasturzio, and M~Amoretti.
\newblock A software tool for mapping and executing distributed quantum
  computations on a network simulator.

\bibitem[Cuomo et~al.(2020)Cuomo, Caleffi, and Cacciapuoti]{Cuomo_2020}
Daniele Cuomo, Marcello Caleffi, and Angela~Sara Cacciapuoti.
\newblock Towards a distributed quantum computing ecosystem.
\newblock \emph{IET Quantum Communication}, 1\penalty0 (1):\penalty0 3–8, Jul
  2020.
\newblock ISSN 2632-8925.
\newblock \doi{10.1049/iet-qtc.2020.0002}.
\newblock URL \url{http://dx.doi.org/10.1049/iet-qtc.2020.0002}.

\bibitem[Gyongyosi and Imre(2021)]{gyongyosi2021distributed}
L~Gyongyosi and S~Imre.
\newblock Distributed quantum computation for near-term quantum environments.
\newblock In \emph{Quantum Information Science, Sensing, and Computation XIII},
  volume 11726, page 117260I. International Society for Optics and Photonics,
  2021.

\bibitem[Preskill(1998)]{preskill1998fault}
John Preskill.
\newblock Fault-tolerant quantum computation.
\newblock In \emph{Introduction to quantum computation and information}, pages
  213--269. World Scientific, 1998.

\bibitem[Shor(1996)]{shor1996fault}
Peter~W Shor.
\newblock Fault-tolerant quantum computation.
\newblock In \emph{Proceedings of 37th Conference on Foundations of Computer
  Science}, pages 56--65. IEEE, 1996.

\bibitem[Shor(1997)]{Shor_1997}
Peter~W. Shor.
\newblock Polynomial-time algorithms for prime factorization and discrete
  logarithms on a quantum computer.
\newblock \emph{SIAM Journal on Computing}, 26\penalty0 (5):\penalty0
  1484–1509, Oct 1997.
\newblock ISSN 1095-7111.
\newblock \doi{10.1137/s0097539795293172}.
\newblock URL \url{http://dx.doi.org/10.1137/S0097539795293172}.

\bibitem[Harrow et~al.(2009)Harrow, Hassidim, and Lloyd]{Harrow_2009}
Aram~W. Harrow, Avinatan Hassidim, and Seth Lloyd.
\newblock Quantum algorithm for linear systems of equations.
\newblock \emph{Physical Review Letters}, 103\penalty0 (15), Oct 2009.
\newblock ISSN 1079-7114.
\newblock \doi{10.1103/physrevlett.103.150502}.
\newblock URL \url{http://dx.doi.org/10.1103/PhysRevLett.103.150502}.

\bibitem[et. al.(2021{\natexlab{a}})]{mi2021observation}
Xiao~Mi et. al.
\newblock Observation of time-crystalline eigenstate order on a quantum
  processor, 2021{\natexlab{a}}.

\bibitem[Arute et~al.(2019)Arute, Arya, Babbush, Bacon, Bardin, Barends,
  Biswas, Boixo, Brandao, Buell, et~al.]{arute2019quantum}
Frank Arute, Kunal Arya, Ryan Babbush, Dave Bacon, Joseph~C Bardin, Rami
  Barends, Rupak Biswas, Sergio Boixo, Fernando~GSL Brandao, David~A Buell,
  et~al.
\newblock Quantum supremacy using a programmable superconducting processor.
\newblock \emph{Nature}, 574\penalty0 (7779):\penalty0 505--510, 2019.

\bibitem[Harrow and Montanaro(2017)]{harrow2017quantum}
Aram~W Harrow and Ashley Montanaro.
\newblock Quantum computational supremacy.
\newblock \emph{Nature}, 549\penalty0 (7671):\penalty0 203--209, 2017.

\bibitem[Bremner et~al.(2017)Bremner, Montanaro, and
  Shepherd]{bremner2017achieving}
Michael~J Bremner, Ashley Montanaro, and Dan~J Shepherd.
\newblock Achieving quantum supremacy with sparse and noisy commuting quantum
  computations.
\newblock \emph{Quantum}, 1:\penalty0 8, 2017.

\bibitem[Yetis and Karakoes(2021)]{9390130}
Hasan Yetis and Mehmet Karakoes.
\newblock Investigation of noise effects for different quantum computing
  architectures in ibm-q at nisq level.
\newblock In \emph{2021 25th International Conference on Information Technology
  (IT)}, pages 1--4, 2021.
\newblock \doi{10.1109/IT51528.2021.9390130}.

\bibitem[Experience()]{ibm-quantum-experience}
IBM~Quantum Experience.
\newblock URL \url{https://quantum-computing.ibm.com/}.

\bibitem[Saleem et~al.(2021)Saleem, Tomesh, Perlin, Gokhale, and
  Suchara]{saleem2021quantum}
Zain~H. Saleem, Teague Tomesh, Michael~A. Perlin, Pranav Gokhale, and Martin
  Suchara.
\newblock Quantum divide and conquer for combinatorial optimization and
  distributed computing, 2021.

\bibitem[Preskill(2015)]{preskill_2015}
John Preskill.
\newblock Ph219/cs219 quantum computation 2018-19, Jul 2015.
\newblock URL \url{http://theory.caltech.edu/~preskill/ph219/ph219_2018-19}.

\bibitem[Tang et~al.(2021)Tang, Tomesh, Suchara, Larson, and
  Martonosi]{tang2021cutqc}
Wei Tang, Teague Tomesh, Martin Suchara, Jeffrey Larson, and Margaret
  Martonosi.
\newblock Cut{QC}: {U}sing small quantum computers for large quantum circuit
  evaluations.
\newblock In \emph{Proceedings of the 26th ACM International Conference on
  Architectural Support for Programming Languages and Operating Systems}, pages
  473--486, 2021.

\bibitem[Peng et~al.(2020)Peng, Harrow, Ozols, and Wu]{peng2020simulating}
Tianyi Peng, Aram~W Harrow, Maris Ozols, and Xiaodi Wu.
\newblock Simulating large quantum circuits on a small quantum computer.
\newblock \emph{Physical Review Letters}, 125\penalty0 (15):\penalty0 150504,
  2020.

\bibitem[Ayral et~al.(2021)Ayral, Régent, Saleem, Alexeev, and
  Suchara]{ayral2020quantum}
Thomas Ayral, François-Marie~Le Régent, Zain Saleem, Yuri Alexeev, and Martin
  Suchara.
\newblock Quantum divide and compute: Exploring the effect of different noise
  sources.
\newblock \emph{SN Computer Science}, 2\penalty0 (3), Mar 2021.
\newblock ISSN 2661-8907.
\newblock \doi{10.1007/s42979-021-00508-9}.
\newblock URL \url{http://dx.doi.org/10.1007/s42979-021-00508-9}.

\bibitem[Siraichi et~al.(2018)Siraichi, Santos, Collange, and
  Pereira]{siraichi2018qubit}
Marcos~Yukio Siraichi, Vin{\'\i}cius Fernandes~dos Santos, Sylvain Collange,
  and Fernando Magno~Quint{\~a}o Pereira.
\newblock Qubit allocation.
\newblock In \emph{Proceedings of the 2018 International Symposium on Code
  Generation and Optimization}, pages 113--125, 2018.

\bibitem[Zulehner et~al.(2018)Zulehner, Paler, and
  Wille]{zulehner2018efficient}
Alwin Zulehner, Alexandru Paler, and Robert Wille.
\newblock An efficient methodology for mapping quantum circuits to the ibm qx
  architectures.
\newblock \emph{IEEE Transactions on Computer-Aided Design of Integrated
  Circuits and Systems}, 38\penalty0 (7):\penalty0 1226--1236, 2018.

\bibitem[Wille et~al.(2019)Wille, Burgholzer, and Zulehner]{wille2019mapping}
Robert Wille, Lukas Burgholzer, and Alwin Zulehner.
\newblock Mapping quantum circuits to ibm qx architectures using the minimal
  number of swap and h operations.
\newblock In \emph{2019 56th ACM/IEEE Design Automation Conference (DAC)},
  pages 1--6. IEEE, 2019.

\bibitem[Nash et~al.(2020)Nash, Gheorghiu, and Mosca]{nash2020quantum}
Beatrice Nash, Vlad Gheorghiu, and Michele Mosca.
\newblock Quantum circuit optimizations for nisq architectures.
\newblock \emph{Quantum Science and Technology}, 5\penalty0 (2):\penalty0
  025010, 2020.

\bibitem[Tan and Cong(2020)]{tan2020optimal}
Bochen Tan and Jason Cong.
\newblock Optimal layout synthesis for quantum computing.
\newblock In \emph{2020 IEEE/ACM International Conference On Computer Aided
  Design (ICCAD)}, pages 1--9. IEEE, 2020.

\bibitem[Cowtan et~al.(2020{\natexlab{a}})Cowtan, Simmons, and
  Duncan]{cowtan2020generic}
Alexander Cowtan, Will Simmons, and Ross Duncan.
\newblock A generic compilation strategy for the unitary coupled cluster
  ansatz, 2020{\natexlab{a}}.

\bibitem[Cowtan et~al.(2020{\natexlab{b}})Cowtan, Dilkes, Duncan, Simmons, and
  Sivarajah]{Cowtan_2020}
Alexander Cowtan, Silas Dilkes, Ross Duncan, Will Simmons, and Seyon Sivarajah.
\newblock Phase gadget synthesis for shallow circuits.
\newblock \emph{Electronic Proceedings in Theoretical Computer Science},
  318:\penalty0 213–228, May 2020{\natexlab{b}}.
\newblock ISSN 2075-2180.
\newblock \doi{10.4204/eptcs.318.13}.
\newblock URL \url{http://dx.doi.org/10.4204/EPTCS.318.13}.

\bibitem[et. al.(2021{\natexlab{b}})]{Qiskit}
MD~SAJID~ANIS et. al.
\newblock Qiskit: An open-source framework for quantum computing,
  2021{\natexlab{b}}.

\bibitem[Bernstein and Vazirani(1997)]{doi:10.1137/S0097539796300921}
Ethan Bernstein and Umesh Vazirani.
\newblock Quantum complexity theory.
\newblock \emph{SIAM Journal on Computing}, 26\penalty0 (5):\penalty0
  1411--1473, 1997.
\newblock \doi{10.1137/S0097539796300921}.
\newblock URL \url{https://doi.org/10.1137/S0097539796300921}.

\bibitem[Perlin et~al.(2021)Perlin, Saleem, Suchara, and
  Osborn]{perlin2021quantum}
Michael~A. Perlin, Zain~H. Saleem, Martin Suchara, and James~C. Osborn.
\newblock Quantum circuit cutting with maximum likelihood tomography, 2021.

\bibitem[Chen et~al.(2014)Chen, Neill, Roushan, Leung, Fang, Barends, Kelly,
  Campbell, Chen, Chiaro, Dunsworth, Jeffrey, Megrant, Mutus, O'Malley,
  Quintana, Sank, Vainsencher, Wenner, White, Geller, Cleland, and
  Martinis]{PhysRevLett.113.220502}
Yu~Chen, C.~Neill, P.~Roushan, N.~Leung, M.~Fang, R.~Barends, J.~Kelly,
  B.~Campbell, Z.~Chen, B.~Chiaro, A.~Dunsworth, E.~Jeffrey, A.~Megrant, J.~Y.
  Mutus, P.~J.~J. O'Malley, C.~M. Quintana, D.~Sank, A.~Vainsencher, J.~Wenner,
  T.~C. White, Michael~R. Geller, A.~N. Cleland, and John~M. Martinis.
\newblock Qubit architecture with high coherence and fast tunable coupling.
\newblock \emph{Phys. Rev. Lett.}, 113:\penalty0 220502, Nov 2014.
\newblock \doi{10.1103/PhysRevLett.113.220502}.
\newblock URL \url{https://link.aps.org/doi/10.1103/PhysRevLett.113.220502}.

\bibitem[Yan et~al.(2018)Yan, Krantz, Sung, Kjaergaard, Campbell, Orlando,
  Gustavsson, and Oliver]{PhysRevApplied.10.054062}
Fei Yan, Philip Krantz, Youngkyu Sung, Morten Kjaergaard, Daniel~L. Campbell,
  Terry~P. Orlando, Simon Gustavsson, and William~D. Oliver.
\newblock Tunable coupling scheme for implementing high-fidelity two-qubit
  gates.
\newblock \emph{Phys. Rev. Applied}, 10:\penalty0 054062, Nov 2018.
\newblock \doi{10.1103/PhysRevApplied.10.054062}.
\newblock URL \url{https://link.aps.org/doi/10.1103/PhysRevApplied.10.054062}.

\bibitem[Mundada et~al.(2019)Mundada, Zhang, Hazard, and
  Houck]{PhysRevApplied.12.054023}
Pranav Mundada, Gengyan Zhang, Thomas Hazard, and Andrew Houck.
\newblock Suppression of qubit crosstalk in a tunable coupling superconducting
  circuit.
\newblock \emph{Phys. Rev. Applied}, 12:\penalty0 054023, Nov 2019.
\newblock \doi{10.1103/PhysRevApplied.12.054023}.
\newblock URL \url{https://link.aps.org/doi/10.1103/PhysRevApplied.12.054023}.

\bibitem[Li et~al.(2020)Li, Cai, Yan, Wang, Pan, Ma, Cai, Han, Hua, Han, Wu,
  Zhang, Wang, Song, Duan, and Sun]{PhysRevApplied.14.024070}
X.~Li, T.~Cai, H.~Yan, Z.~Wang, X.~Pan, Y.~Ma, W.~Cai, J.~Han, Z.~Hua, X.~Han,
  Y.~Wu, H.~Zhang, H.~Wang, Yipu Song, Luming Duan, and Luyan Sun.
\newblock Tunable coupler for realizing a controlled-phase gate with
  dynamically decoupled regime in a superconducting circuit.
\newblock \emph{Phys. Rev. Applied}, 14:\penalty0 024070, Aug 2020.
\newblock \doi{10.1103/PhysRevApplied.14.024070}.
\newblock URL \url{https://link.aps.org/doi/10.1103/PhysRevApplied.14.024070}.

\bibitem[Han et~al.(2020)Han, Cai, Li, Wu, Ma, Ma, Wang, Zhang, Song, and
  Duan]{PhysRevA.102.022619}
X.~Y. Han, T.~Q. Cai, X.~G. Li, Y.~K. Wu, Y.~W. Ma, Y.~L. Ma, J.~H. Wang, H.~Y.
  Zhang, Y.~P. Song, and L.~M. Duan.
\newblock Error analysis in suppression of unwanted qubit interactions for a
  parametric gate in a tunable superconducting circuit.
\newblock \emph{Phys. Rev. A}, 102:\penalty0 022619, Aug 2020.
\newblock \doi{10.1103/PhysRevA.102.022619}.
\newblock URL \url{https://link.aps.org/doi/10.1103/PhysRevA.102.022619}.

\end{thebibliography}


\end{document}